\documentstyle[12pt]{article}
\normalsize
\def\sp{~~~~~}

\def\a{\alpha}
\def\b{\beta}

\def\e{\epsilon}

\def\g{\gamma}
\def\h{\eta}

\def\m{\mu}
\def\n{\nu}

\def\p{\pi}
\def\q{\theta}

\def\s{\sigma}

\def\bar#1{\overline{#1}}

\def\Hat#1{\rlap{\kern.10em$\widehat{\phantom G}$}#1}
\def\HAt#1{\rlap{\kern.05em$\widehat{\phantom G}$}#1}

\def\cap#1{\rlap{\kern.1em$\widehat{\phantom{G\vrule height.8em}}$}#1{}}
\def\Cap#1{\rlap{\kern.05em$\widehat{\phantom{G\vrule height.8em}}$}#1{}}

\let\oldtheequation=\theequation
\def\doteqs#1{\setcounter{equation}{0}
            \def\theequation{{#1}.\oldtheequation}}

\newcounter{sxn}
\def\sx#1{\addtocounter{sxn}{1} \bigskip\medskip \goodbreak \noindent{\large\bf
\centerline{\thesxn.~~#1}} \nobreak \medskip}
\def\sxn#1{\sx{#1} \doteqs{\thesxn}}

\newcounter{axn}

\def\br{\begin{thebibliography}{abc}}
\def\er{\end{thebibliography}}

\date{}

\begin{document}
\bibliographystyle{unsrt}
\footskip 1.0cm
\thispagestyle{empty}
\begin{center} {\Large \bf Theoretical Physics Institute\\
University of Minnesota}\end{center}
\begin{flushright}
\vspace*{4mm}
TPI-MINN-91/38-T\\
SU-4228-492\\
October 1991\\
\end{flushright}
\vspace*{6mm}
\centerline {\LARGE CHERN-SIMONS DYNAMICS}
\vspace*{3mm}
\centerline {\LARGE AND THE QUANTUM HALL EFFECT \footnote {\small To be
published in a volume in honour of Professor R. Vijayaraghavan.} }
\vspace*{10mm}
\centerline {\large A. P. Balachandran}
\vspace*{5mm}
\centerline {\it Department of Physics, Syracuse University,}
\centerline {\it Syracuse, NY 13244-1130}
\vspace*{10mm}
\centerline {\large A.M. Srivastava}
\vspace*{5mm}
\centerline {\it Theoretical Physics Institute, University of Minnesota}
\centerline {\it Minneapolis, Minnesota 55455}

\vspace*{10mm}

\centerline {\bf ABSTRACT}

 Theoretical developments during the past several years have shown that large
scale properties of the Quantum Hall system can be successfully described by
effective field theories which use the Chern-Simons interaction.  In this
article, we first recall certain salient features of the Quantum Hall Effect
and their microscopic explanation.  We then review one particular approach to
their description based on the Chern-Simons Lagrangian and its variants.

\newpage

\newcommand{\be}{\begin{equation}}
\newcommand{\ee}{\end{equation}}

\baselineskip=24pt
\setcounter{page}{1}
\sxn{INTRODUCTION}

 Since the discovery of the Quantum Hall Effect (QHE) in 1980, there have
been significant developments in its theoretical as well as experimental
investigations $^{(1,2)}$.

  QHE was observed in effectively two dimensional systems of electrons
(experimentally realized in terms of inversion layers formed at the
interface between a semiconductor and an insulator or between two
semiconductors) subjected to strong magnetic fields. For such a system, the
classical Hall conductivity is given by

 $$ \sigma_H = {nec \over B} \eqno(1.1)$$

\noindent where $n$ is the electron concentration, $ e = - |e|$ is the charge
of the electron  and $B$ is the magnetic field
perpendicular to the plane of the system.

 However it was observed that at very low temperatures instead of linearly
rising with $n/B$, the Hall conductivity becomes quantized and develops a
series of plateaus given by

 $$ \sigma_H = - {\nu e^2 \over h} \eqno(1.2)$$

\noindent where $\nu$ is an integer.

 Theoretical understanding of this Integer Quantum Hall Effect (IQHE) was
provided in terms of a noninteracting electron system. States of a two
dimensional electron system (without any impurities) in a magnetic field
normal to the two dimensional surface are discrete Landau levels. We can
define the filling factor of Landau levels as $\nu = {n \over n_B}$ where
$n_B = {1 \over 2\pi l^2}$ is the number of states per unit area of a
Landau level ($l$ here being the magnetic length,
$l = (\hbar c/eB)^{1/2}$). Due to
gaps in the single particle density of states, diagonal resistance vanishes
when a Landau level is full and the Fermi level lies in the gap between
occupied levels. Presence of impurities broadens Landau levels and leads
to the presence of localized states in the energy gaps. These localized
states can not carry any current. Therefore increase in the occupation of
these states does not change Hall conductivity. As long as the extended
states in the $\nu^{th}$
Landau level are completely occupied and it is only the localized states (lying
in between $\nu^{th}$ and $\nu +1^{st}$ Landau levels) which are being
further occupied as $n/B$ is increased, the Hall conductivity will remain
constant at a value given by Eq.(1.1) with $n = \nu n_B$. This gives the result
of Eq.(1.2) explaining the plateaus. Longitudinal resistance becomes
nonzero and the Hall conductivity makes a transition from one plateau to
the next when extended states in the next Landau level start
getting filled.

 Soon after the discovery of IQHE, it was discovered in 1982 that in systems
with extremely low impurity concentrations and at very low temperatures, the
value of $\nu$ in Eq.(1.2) can assume certain rational fractional values f.
This effect is known as the Fractional Quantum Hall Effect (FQHE). It was
further observed that if f = p/q where the integers p and q have no common
factor, then q was necessarily odd. [We may point out that some later
observations also revealed even denominator values of f. In terms of the
Laughlin theory of FQHE  to be described below, these even denominators
are understood as arising from an electron system where electrons are bound
in pairs and the pairs behave like bosons (cf. ref. 2) or alternatively are
``bound" to magnetic vortices in a sense we indicate in Section 3.]

 Although FQHE and IQHE seem very similar, the theoretical understanding of
FQHE required radically new concepts. While the theory of IQHE was based on
a model of a noninteracting system of electrons, the theory of FQHE utilizes
strong correlations among electrons. Laughlin proposed that the ground state
of the electron system in FQHE is a translationally invariant liquid state and
that the lowest energy charge excitations in the system are fractionally
charged quasiparticles and quasiholes. These excitations obey fractional
statistics and therefore are anyons. Hall plateaus observed in FQHE experiments
are then explained as being due to localization of these fractionally charged
quasiparticles. Quantized Hall conductivity is still given by Eq.(1.2),
but now the charge carriers are fractionally charged quasiparticles leading
to fractional values of the filling factor $\nu$ in Eq.(1.2) as we will
explain below.

 Let us describe the constant imposed magnetic field $\vec B = (0,0,B)$ in
a gauge where the vector potential $\vec A$ is 1/2 ($\vec B \times \vec x$),
$\vec x = (x,y,0)$ being a point on the two-dimensional surface. The wave
function proposed by Laughlin for FQHE corresponding to the $\nu = {1 \over m}$
ground state is, in this gauge,

 $$\Psi_m = \prod_{j<k}^{N} (z_j - z_k)^m \prod_{j=1}^{N}
e^{-|z_j|^2/4l^2} \eqno(1.3)$$

\noindent where $N$ is number of electrons, $z_j = x_j + iy_j$ is the position
of the $j^{th}$ electron in complex coordinates and $l$ is the magnetic length
introduced earlier. $m$ is an odd integer so that $\Psi_m$ is antisymmetric
in $z_j$ as required by the Fermi statistics of electrons.

  The probability density of electrons in state $\Psi_m$ can be
written as

 $$ |\Psi_m|^2 = e^{-H_m}$$

\noindent where

 $$H_m = -2m \sum_{j<k} ln|z_j - z_k| + \sum_{j} |z_j|^2/2l^2 \eqno(1.4)$$

  We note that $H_m$ can be identified with the potential energy of a
two-dimensional, one component plasma where particles of charge $m$ repel
one another via logarithmic interaction and are attracted to the origin by
a uniform neutralizing background charge density $\rho = {1 \over 2\pi l^2}$.
Charge neutrality of the plasma will then be achieved when the electron
density is equal to 1/$m$ times the charge density of the equivalent
plasma and thus equal to

  $$ \rho_m = {1 \over 2\pi m l^2} \eqno(1.5)$$

  This therefore will give rise to plateaus in the Hall conductivity at
filling factors $\nu = 1/m$. At high densities, $m$ = 1 (full  Landau
level) is energetically most favorable and as the neutralizing background
density of the original electron system is decreased, first the $m$ = 3 state
becomes stable and then $m$ = 5 etc.

 Starting from a ground state as in
Eq.(1.3), one can easily establish that quasiparticle and quasihole
excitations of state $\Psi_m$ are fractionally charged particles obeying
fractional statistics and are therefore anyons. The states with $m$ = 1, 3,
5, ... etc. correspond to filling factors $\nu$ = 1, 1/3, 1/5, ... and are
called parent states. The whole hierarchy of Fractional Quantum Hall
(FQH) states can be generated by essentially repeating the whole process
of construction of the state $\Psi_m$ for the quasiparticles. Thus we
consider a system of quasiparticles condensing into a FQH state of the
form given in Eq.(1.3) where $m$ should be replaced by an appropriate
real number consistent with the statistical properties of quasiparticles.
This process can be iterated for higher levels of the hierarchy.

 Though there are important differences between IQHE and FQHE in
experimental as well as theoretical aspects, there are strong
similarities as well underlying these Quantum Hall (QH) systems.
For example Hall conductivity (and Hall conductance) do not depend
on microscopic details (such as frequencies etc.) or on the geometrical
shape of the sample. These aspects of QH systems are similar to the
ones in the theory of critical phenomena in statistical mechanics
where certain quantities like critical exponents characterizing
continuous phase transitions depend only on the large scale properties
of the underlying statistical system and not on the microscopic
dynamics. Motivated by this universal behavior of QH systems, many
authors$^{(3,4)}$ have developed effective field theories for the large scale
behavior of QH systems. The work of Fr\"{o}hlich and Kerler, and Fr\"{o}hlich
and Zee$^{(4)}$ in this regard is of particular interest for the present
article. They argued that large scale properties of QH systems can be
described in terms of a pure abelian Chern-Simons theory where the
Hall conductivity turns out to be inversely proportional to the
coefficient of the Chern-Simons action.

 In the rest of the article, we will elaborate on the theoretical
developments centered around the pure Chern-Simons description of
QH systems. One of the aspects of these systems which plays a very
important role in understanding their large scale behavior is the existence
of edge currents$^{(5)}$. Therefore, in Section 2, we first describe the origin
of these edge currents from the microscopic physics of a Quantum Hall
system. In Section 3, we will discuss the relation between the theory
of the QH system and Chern-Simons gauge theory following the papers
of Fr\"{o}hlich and Kerler, and Fr\"{o}hlich and Zee$^{(4)}$ and show how
Chern-
Simons theory leads to fractional quantization of $\sigma_H$. In Section
4, we will show how the edge currents of a Quantum Hall system can be
obtained from a Chern-Simons theory following the approach of
Balachandran et al.$^{(6)}$. [The existence of edge states in a Chern-
Simons theory is first due to Witten$^{(7)}$.]

\sxn{MASSLESS EDGE CURRENTS IN QUANTUM  HALL SYSTEMS}

 It was pointed out by Halperin$^{(5)}$ that in a QH system, there are
current carrying edge states which extend along the perimeter of the
system. Following the discussion in ref. 5, let us consider a system
of electrons on a two-dimensional plane with annular geometry as shown
in Fig. 1. There is a uniform magnetic field B through the annulus
perpendicular to the plane and in addition there is a flux $\Phi$
going through the hole ($r < r_1$). [Here $r = [(x^1)^2 + (x^2)^2]^{1/2},
(x^1, x^2)$ being the coordinates of the plane. We have slightly
changed notations from those in Section 1 for later convenience.] We work
in a gauge where $A_r$ = 0 and

  $$A_\theta = {1 \over 2} B r  + {\Phi \over 2\pi r}, \eqno(2.1)$$

\noindent $A_r$ and $A_\theta$ being the radial and azimuthal components
of the vector potential $A$.

  Due to the azimuthal symmetry, the third component of orbital angular
momentum is a good quantum number and the states of electrons
in the interior of the  annulus (at distances from the edges large compared
to the magnetic length $l$) are given by the Landau states

 $$\Psi_{m,\nu}(\vec r) = const. \times e^{im\theta} f_\nu(r - r_m).
\eqno(2.2)$$

\noindent Here $m$ and $\nu$ are integers, $m$ being the magnetic quantum
number, and $f_\nu$ is the $\nu + 1^{st}$ eigenstate of a one dimensional
shifted harmonic oscillator with center $r_m$ given by

 $$B \pi r_m^2 = m \Phi_0 - \Phi,$$

 $$ \Phi_0 = {hc \over e} = {\rm Flux \ quantum}. \eqno(2.3)$$

\noindent  $f_\nu$ is localized near $r_m$, decreasing exponentially away from
$r_m$ with typical scale $l$. The energy of the Landau state $\Psi_{m,\nu}$ is

 $$ E_{m,\nu} = \hbar \omega_c (\nu + {1 \over 2}) \eqno(2.4)$$

\noindent where $\omega_c = |eB| / {\rm m}^*c$ is the cyclotron frequency,
m$^*$ being the effective mass of the electrons. The current
carried by state $\Psi_{m,\nu}$ is

 $$I_{m,\nu} = {e \over {\rm m}^*} \int_{0}^{\infty} dr |\Psi_{m,\nu}(\vec
r)|^2
 [{m \hbar \over r} - { e A(r) \over c}] $$

 \noindent where the integration over $r$ is performed at a fixed $\theta$.
On using Eq.(2.1) and (2.3) and the fact that $|\Psi_{m,\nu}(\vec r)|^2$
is symmetric about $r_m$ decreasing exponentially
away from $r_m$, it becomes

 $$I_{m,\nu} \simeq {e^2 B_0 \over {\rm m}^* c} \int_{0}^{\infty} dr
|\Psi_{m,\nu}(\vec r)|^2 (r_m - r).\eqno(2.5) $$

 Now we note that since $|\Psi_{m,\nu}|^2$ is symmetric about $r = r_m$
for the interior of the annulus, the integral in Eq. (2.5) effectively
vanishes when $r_m$ is far from the edges of the annulus. However, when
$r_m$ is close to the edges of the annulus (closer than a few times $l$), the
boundary condition that wave functions vanish at the edges makes
$|\Psi_{m,\nu}|^2$ asymmetric about $r_m$ in that region and the integral
need not vanish. There are thus currents at the edges of the sample. The
effect of boundary conditions at the edges is to shift the energy levels as
shown in Fig. 2.

 The existence of these edge currents can be easily demonstrated using very
general arguments as well. In the absence of externally applied potential
differences, if the electrons are confined in a two dimensional surface
with boundaries, then there must be potential barriers at the boundaries
so that electrons do not escape. The gradient of such a potential near an edge
will give rise to a force which acts like an electric field $\vec E$ directed
radially outwards and whose net effect is to confine the electrons. One
can think of this electric field as arising from an accumulation of a net
positive charge near the edge of the region which the electron is trying
to escape. When a magnetic field $\vec B$, perpendicular to the plane
of the sample, is also present, this $\vec E$ gives rise to a Hall current in
the direction $\vec E \times \vec B$ which is tangent to the edge. This
current is hence confined to the edge.
The potential barrier at the edge will be expected to change $E_{m,\nu}$
as shown in Fig.2. As mentioned above, this is indeed also what happens to
the energy levels as a consequence of the boundary conditions on wave
functions requiring them to vanish at the edges.

 Using Eq.(2.1) and (2.3), we can get the following expression for the
edge currents:

 $$I_{m,\nu} = -c {\partial E_{m,\nu} \over \partial \Phi} = {e \over h}
{\partial E_{m,\nu} \over \partial m}. \eqno(2.6)$$

\noindent From this we note that the currents have opposite directions near the
outer and inner edges of the annulus. It was further shown by Halperin that
a moderate amount of disorder does not destroy these edge currents.

 We thus find theoretically that when $\nu$ Landau levels are filled
(the filling factor is $\nu$), there are $\nu$ current carrying states at
each of the edges of the sample. Further, the edge currents at the two edges
flow in opposite directions. These conclusions have been confirmed in
numerical work as well by Rammal et al. (see the article by Prange in the first
book of ref. 1),
who find that for filling factor $\nu$, there are $\nu$ pairs of eigenstates
localized at the two edges with opposite momenta.

 In this discussion, we have neglected the spin degree of freedom of
the electrons, this neglect being justified if the magnetic field
is strong. The edge excitations for strong magnetic fields are
therefore scalar, chiral fermions propagating in one dimension along
the edges of the sample. These fermions will approximately have the
dynamics of a ``relativistic" massless particle with the one dimensional
momentum given by

 $$ k = {m - m_{1,2}^F \over r_{1,2}} \eqno(2.7)$$

\noindent and speed $k/E_{m,\nu}$, this speed being the substitute for the
speed of light for these particles. Here $m_1^F$ and $m_2^F$ are the magnetic
quantum numbers corresponding to filled Fermi levels at the inner and outer
edges of the sample respectively. [``Relativistic" kinematics enters this
problem for certain standard reasons: The energy $E$ of an edge state as
measured from the Fermi energy $E^F$ is, to leading order, proportional
to $k$.]

 This concludes the demonstration of the existence of massless edge
currents in QH systems using microscopic arguments. In the next two Sections,
we explore the universal aspects of QH systems for large scale observations
and outline their description using the Chern-Simons gauge theory.

\sxn{RELATION OF QHE TO CHERN-SIMONS GAUGE THEORY}

In this Section we will review certain results due to Fr\"{o}hlich and Kerler,
and Fr\"{o}hlich and Zee$^{(4)}$ who show that the QH System is related to pure
Chern
Simons gauge theory and to certain rational conformal field theories.  Such
relations could have been anticipated if it is recalled  that there are chiral
edge currents in a QH system and that according to Witten$^{(7)}$,
Chern-Simons theory on a two-dimensional space is equivalent to a chiral
current
algebra on the boundary of that space. One can thus guess the existence of a
correspondence between Chern-Simons theory and the QH system.

 We will set the speed of light $c$ equal to 1 in this Section so that
magnetic flux can be measured in units of $h/e$.

Let us begin our discussion by examining a QH system characterized by zero
longitudinal resistance.  The conductivity tensor $\sigma$ can then be written
as

\be
\s = \left ( \begin{array}{cc}
0 & \s_H \\
-\s_H & 0
\end{array} \right )
\ee

 In QH systems, $\s_{H}$ is quantized and is a rational multiple of
$e^{2}/h$.  The idea pursued in ref. 4 is that this fact may have a
universal explanation emerging from rational conformal field theories.

As the longitudinal conductivity $\s_{L}$ is zero for a two-dimensional
system with $\s$ given by Eq. (3.1), the current density $j$ induced by an
electric field $E$ is given by
\be
j^{\a}(\vec{x},t)=\s_{H} \e^{\a \b} E_{\b}(\vec{x},t);~~~\a , \b =1,2; ~
\e^{\a\b}=-\e^{\b \a}, \e^{12}=1\ .
\ee
Here $E_{\a} = -F_{0\a},  F_{\m \n}$ being the electromagnetic
field strength tensor.

Now if $j^0$ is the charge density, then we have the continuity equation
\be
\frac {\partial j^0}{\partial x^0} + {\vec \bigtriangledown}\cdot
\vec {j} = 0,~~~x^0=t \ .
\ee
Also $B$ and $E$ are related by the Maxwell's equation
\be
\frac {\partial B}{\partial x^0} = - \e^{\a \b} \partial_{\a} E_{\b}\ .
\ee

\noindent where $B = F_{12}$. Equations (3.2), (3.3) and (3.4) give
\be
\s_H \frac {\partial B} {\partial x^0}
=  \frac {\partial}{\partial x^0} j^0 \ .
\ee

We thus obtain
\be
j^0 =  \s_H (B+B_c) \ ,
\ee

\noindent Here $B_c$ is an integration constant representing a time independent
background magnetic field.

 Let us assume that the three-dimensional manifold $M$ has the topology
of ${\bf R}^1 \times D$ with $D$ characterizing the two-dimensional
space of the sample, and ${\bf R}^1$ describing time. Furthermore, let $\eta =
(\eta_{\mu \nu})$ be any metric of Euclidean or Lorentzian signature on $M$.
Then Eqs. (3.2) and (3.6) can be extended to a
generally covariant form valid for arbitrary metrics as well as
follows.

Let
\be
J_{\a \b}(x)=|Det \ \h (x)|^{-1/2} \e_{\a \b \g} j^{\g}(x), ~~x=
\vec {x},t
\ee
and
\be
j^{\a}(x) = {1 \over 2} |Det \ \h (x)|^{1/2} \s_H \e^{\a \b \g} F_{\b \g}(x) \
,
\ee
\noindent
Here $\e_{\a \b \g}$ is the totally antisymmetric symbol with $\e_{012}=1$ and
$t=x^0$ is time.  Then
\be
J_{\a \b}(x) = \s_H F_{\a \b}(x)
\ee
\noindent
(3.9) reduces to (3.2) and (3.6) for a flat metric.

Using the language of differential forms, we can write Eqs. (3.9) and (3.7) as
\be
J=\s_H F \ ;
\ee
\be
J=^*j
\ee
where $J = {1 \over 2} J_{\a \b} dx^{\a} \wedge dx^{\b}$ and * is the
Hodge dual. The one form $j(x)$ is defined as
$$j(x) = \sum_{\a} \left (\sum_{\b} \h_{\a \b}(x) j^{\b}(x)\right )
dx^{\a}\ . $$
\noindent
The continuity equation (3.3) can be written as
\be
d J = 0
\ee
where $d$ is the exterior derivative.

 We shall assume that $\s_H$ is a constant. Equation (3.10) then gives
the Maxwell equations
\be
dF = 0
\ee
Here, we can write $F=dA^{\prime},~A^{\prime}=A+A_c$ where $A_c$ is the vector
potential corresponding to a constant magnetic field $B_c$ (see Eq. 3.6), $A$
represents the vector potential of a fluctuation field due to
localized sources and $A^{\prime}$ the total vector potential.

Now, Eq. (3.12) implies that
\be
J=da
\ee
where $a$ is a one form.  Equation (3.10) can then be written in terms of one
forms
$a$ and $A^{\prime}$ as

 $$da = \s_H dA^{\prime} $$

We now note that this last equation can be obtained from an action principle
with the action $S_{CS}$ given by
\be
S_{CS} = \frac{1}{2\s_{H}} \int_M (a-\s_H A^{\prime})\wedge d
(a-\s_HA^{\prime})
\ee
\noindent
or in terms of components,

 $$S_{CS} = \frac {1}{2\s_{H}} \int_M \e^{\a \b \g}(a_\a-\s_HA^{\prime}_{\a})
\partial_\b (a_{\g}-\s_H A^{\prime}_{\g})d^3x . \ $$

\noindent The overall normalization of $S_{CS}$ is here fixed by the
requirement that the coupling of $A^{\prime}_{\mu}$ to  $j^{\mu}$ is by
the term  $-j^{\mu}A^{\prime}_{\mu}$ in the Lagrangian density.

The action $S_{CS}$ is the Chern-Simons action for the gauge field
$a-\s_HA^{\prime}$.

It is important to note at this step that the derivation of Eq. (3.15) from the
QHE is valid only in the scaling limit when both length and 1/frequency scales
are large.  This is because although the continuity equation (3.12) is exact,
Eq.
(3.2) is experimentally observed to be valid only at large distance and time
scales.

The action $S_{CS}$ can be naturally generalized to the case where there are
several independently conserved electric current densities
$j^{(i)},i=1,...m$.  For
example, for $m$ filled Landau levels, if one neglects mixing of levels (which
is a good approximation due to the large gaps between Landau levels), each
level can be treated as dynamically independent with electric currents in each
level being separately conserved.  Then one has
$$J^{(i)} = \s^{(i)}_H F$$
and
\be
J^{(i)} = da^{(i)}\ .
\ee
\noindent
The action in this case is given by
\be
S_{CS}(\{a^{(i)}\},A)=\int_M \sum_{i=1}^{m} \frac {1}{2\s^{(i)}_{H}}
(a^{(i)}-\s^{(i)}_H A^{\prime}) \wedge d (a^{(i)}-\s^{(i)}_H A^{\prime}) \ .
\ee

Let us look back at Eq. (3.15). Recall that $A^{\prime}=A+A_c$ where $A_c$ is
the vector potential of the background magnetic field $B_c$.  Let us introduce
a change of variable by setting $a = {\bar a} + \s_H A_c$ and call $\bar a$
again as $a$.
Then the field equations from Eq. (3.15) relate $a$ and $A$:
\be
da = \s_H dA .
\ee

 This equation implies that a vortex of magnetic flux
$$\Phi = \int  dA$$
\noindent
carries a charge
\be
 q = \s_H \Phi \ .
\ee

 The statistics obeyed by these quasiparticles can be determined in the
following manner.  Consider two such quasiparticles each carrying magnetic
flux $\Phi$ and electric charge $q=\s_H \Phi$.  Under an exchange of
two quasiparticles, the wave function picks up a phase factor $\exp (2\pi i
\q)$
where $\q$ characterizes the quasiparticle statistics.  $\q$ can be calculated
by performing two successive exchanges (which amount to taking one
quasiparticle a full circle around the other quasiparticle) and calculating the
Aharonov-Bohm phase factor. One gets

 $$ {\rm exp} \left [2\pi i \q \right ] =
{\rm exp} \left [- \frac {i}{2\hbar}\ q \Phi \right] $$
 $$ \qquad \qquad ={\rm exp} \left [- \frac {i}{2\hbar}\ \s_H (\Phi)^2 \right]
$$
 $$ \qquad \quad ={\rm exp} \left [- \frac {i\pi {\tilde \Phi}^2}{k} \right] $$

\noindent where we have set
$$\s_H = \frac {e^2}{h k} , $$
\be
{\tilde \Phi} = \frac {e}{h} \Phi . \
\ee

 We thus find,

\be
\q = - \frac {{\tilde \Phi}^{2}}{2 k} {\it mod} N,~~N \in {\bf Z} \ .
\ee
$\q = 0$ and $- \frac {1}{2} ~ (mod N)$ respectively correspond to Bose and
Fermi statistics whereas the quasiparticles are anyons
when $\exp (2 \pi i \q)\neq \pm1$.

 Clearly, the electron must be among the charged excitations of the theory.
Since the magnetic field $B_c$ is strong for us, the electron spin is frozen in
the direction of $B_c$.  The symmetry of the many electron spin wave function
implies that the values $\q_e$ and ${\tilde \Phi}_e$ of
$\q$ and ${\tilde \Phi}$ for electrons are related by

$$\q_e = \frac {1}{2} ~ {\it mod} N = \frac {{\tilde \Phi}^2_e}{2 k}
{\it mod} N,~~N \in {\bf Z} $$

\noindent which gives
\be
{\tilde \Phi}_e^2/k=\pm (2 \ell+1), ~ \ell =0,1,2,... \ .
\ee
Equation (3.19) shows that the charge of this particle is given by
\be
q_e = e \frac {{\tilde \Phi}_e}{k} \ .
\ee
Since we want $q_e= e$, Eqs. (3.22) and (3.23) show that

 $$k= \pm (2\ell +1) . $$

With this value of $k$, one can obtain $\s_H$ from Eq. (3.20):
\be
\s_H = \pm \frac{1}{2 \ell +1} \frac {e^2}{h}, \ell = 0,1,2...\ .
\ee
Equation (3.24) gives the fractionally quantized Hall conductivities
corresponding to the parent states in Laughlin's theory.

 Let us recall from Section 1 that there are experiments where even integer
values of $k$ have been obtained (cf. ref. 2).  These even denominators in
$\nu$ arise naturally in the framework of the present model in the following
manner.  Suppose that the external magnetic field $B_c$ is not very large.
Then as
long as the temperature of the system is not extremely low, the spins of the
electrons will not be completely frozen in the direction of $B_c$ and the
N-electron spin wave function need not be completely symmetric.  For example,
it
could be in a singlet state.  The electron may then appear as a compound state
of
a magnetic vortex (with flux $-\frac {2 h}{e}$ and charge $-e$)
and of a neutral fermion with
spin $\frac {1}{2}$.  Allowed statistics $\q$ for this compound state will then
be, from Eq. (3.21), $\q=0 ~ {\it mod} N, ~N \in {\bf Z} $.
{}From Eq. (3.23) and with the above value of
flux for this picture of the electron, we get $k=2$ and $\s_H=\frac {e^2}{2
h}$, thereby obtaining an even denominator plateau.

We will now generalize the theory in order to get the higher levels of
hierarchies for $\s_H$.  Let us go back to the case of $m$ filled Landau levels
with the Lagrangian given in Eq. (3.17).  From the example of
one filled Landau level, with Eq. (3.20) defining $k$, we know that $k$
characterizes the level of the hierarchy and is hence related to the
filling factor, $k$ being 1 for one completely filled
Landau level.  Suppose now that we consider
$m$ filled Landau levels, with $k=1$ for
each level.  The Lagrangian ${\cal L}$ will be given by Eq. (3.17) with
$\s^{(i)}_H$ set equal to ${e^2 \over h}$ for all $i$ (see Eq. (3.20))
and $a$ redefined as indicated earlier in order to get rid of $A_c$.

In obtaining (3.17), we had neglected interactions between the electrons in the
different Landau levels.  Such interactions will lead to mixing between
different levels.  However, by general arguments we expect that in the scaling
limit, the dominant contributions from such interactions can come only from
dimension three (Chern-Simons like) terms.  If we now assume that the
``interaction'' to be added to $\cal L$ should only involve the total
electromagnetic current
$$J= \sum_i J^{(i)} = \sum da^{(i)}$$
and not say, just one $J^{(i)}$, then the electron-electron
interaction changes the part of the Lagrangian in Eq. (3.17) not
involving $A_\m$ to the form
\be
{\cal L}^{(1)}= {h \over 2e^2} (\sum a^{(i)} \wedge d a^{(i)} +
p (\sum_i a^{(i)})(\sum_i da^{(i)}))
\ee
where $p$ is some real constant.  Although the physical basis of this
assumption
is not clear to us, we shall accept it and proceed to study (3.25).

The Lagrangian (3.25) can be written in the more compact form
\be
{\cal L}^{(1)}= {h \over 2e^2}  a^T{\cal K} da
\ee
by introducing a matrix ${\cal K}=I+ p C$ where $C$ is the $m\times m$ matrix
with each entry equal to 1.  $a$ here is the column with entries $a^{(i)}$

 The field equations following from (3.26) and (3.17) are
\be
{\cal K} da^{(i)} = \frac {e^2}{h} dA . \
\ee
\noindent For an applied field $dA$ (independent of $i$), it shows that

$$ \sum_i da^{(i)} = \frac {e^2}{h} \sum_{i,j} {\cal K}_{i,j}^{-1} dA $$

\noindent and hence that the Hall conductivity is
\be
\s_H = \frac {e^2}{h} \sum_{i,j} {\cal K}_{i,j}^{-1} .
\ee

If we characterize a quasiparticle by a vorticity vector

 $$\Phi = \left ( \begin{array}{c}
               \Phi_1 \\
               \cdot \\
               \cdot \\
                \cdot \\
                \Phi_m
        \end{array} \right ) \nonumber $$

\noindent where $\Phi_i$ is its flux associated with the  $i^{th}$ Landau
level, then (3.27)
shows that the electric charge is described by the charge vector

 $$q = \left ( \begin{array}{c}
               q_1 \\
               \cdot \\
               \cdot \\
                \cdot \\
                 q_m
        \end{array} \right ) \nonumber$$

\noindent where
\be
q_i=  \frac {e^2}{h} \sum_j ({\cal K}^{-1})_{ij}\Phi_j \ .
\ee

 We can repeat the steps used in getting (3.21) for this case.  Thus the
statistics phase $\q$ for this case is
\be
\q = - \frac {1}{2h} (\sum_i q_i \Phi_i)~ {\it mod} N,~~N \in {\bf Z} \
\ee
where again $\q$ is defined by the phase factor $e^{2\p i \q}$ which the wave
function picks up under an exchange of two quasiparticles.

  Now since  $C^{2}=mC$, we have,

 $${\cal K}^{-1}=I-\left ( \frac {p}{1+mp} C \right ) \ .$$

Using this in Eqs. (3.28), (3.29) and (3.30) we get

$$ \s_H =  {e^2 \over h} \frac {m}{1+m p},~~q_i= \frac {e^2}{h} \left ( \Phi_i-
\frac {p}{1+m p} \sum_{j=1}^{m} \Phi_j \right ) $$
and
\be
\q = - \frac {1}{2} {\left ( \frac {e}{h} \right ) }^2
\left [ \sum_i \Phi_i^2 -
\frac {p}{(1+m p)} (\sum \Phi_i)^2 \right ]  {\it mod} N,~~N \in {\bf Z} . \
\ee
For $ \frac {e \Phi_i}{h} = 1, i=1,...m$,  we therefore find,
\be
\s_H = \frac {e^2}{h} \frac {m}{1+m p},~q_i = e \frac {1}{1+mp}~ {\rm and}~
\q = - \frac{1}{2} \left (\frac {m}{1+m p} \right )
{\it mod} N,~~N \in {\bf Z}\ .
\ee
For the simplest case of a single Landau level, $m=1$ and  (3.32) becomes
\be
\s_H = \frac {e^2}{h} \frac {1}{p+1}, q = e \frac {1}{p+1}
{}~{\rm and}~\q= - \frac {1}{2(p+1)}{\it mod} N,~~N \in {\bf Z} \ .
\ee

 The expressions for $\s_H, q_i$ and $\q$ in Eq. (3.31) with  $m=1$
agree with those  obtained earlier in this Section for the single filled
Landau level case (Eqs. (3.19), (3.20) and (3.21)).
One can therefore repeat the type of
arguments used for that case and conclude that $(p+1)$ must be an odd integer
giving us the  odd denominator fractional Hall conductivity.  Note that
Eq. (3.32) gives more general values of fractionally quantized Hall
conductivities than those found for $m$ = 1. Fr\"{o}hlich and Zee$^{(4)}$
discuss further generalizations of (3.25) leading to even more general
possibilities for the fractional Hall conductivity.

We will continue in the next Section with the exploration of the relationship
between the Quantum Hall system and the Chern-Simons theory.  We will
demonstrate how the edge currents (see Section 2) in a Quantum Hall system
arise naturally from the Chern-Simons theory.  This result is first due to
Witten$^{(7)}$. We follow the approach of Balachandran et al$^{(6)}$ who
derive further results in Chern-Simons theory using this
approach.

\sxn{CONFORMAL EDGE CURRENTS}

In this final Section, we develop elementary canonical methods for the
quantization of the abelian Chern-Simons action (considered earlier) on a disc
and show that it predicts the edge currents.  They are in fact described by the
edge states of  Witten carrying a representation of the Kac-Moody$^{(8)}$
algebra. The canonical expression for the generators of diffeomorphisms
(diffeos) on the boundary
of the disk are also found and it is established that they are the Chern-Simons
version of the Sugawara construction.

 The Lagrangians considered here follow from (3.15) by setting

$${\bar a} = (a - \s_H A^{\prime})[2\pi / |k\s_H|]^{1/2}$$
\noindent and calling $\bar a$  again as $a$, $k$ being
$|k|({|\s_H| \over \s_H})$.
We do so in order to be consistent with the form of the
Chern-Simons Lagrangian most frequently encountered in the
literature.

 In this Section, we will use natural units where $\hbar = c = 1$.

\vskip .3in
\centerline {\bf 4.1 THE CANONICAL FORMALISM}
\vskip .3in

\indent
Let us start with a U(1) Chern-Simons (CS) theory on the solid cylinder
$D \times R^{1}$ with action given by
\begin{equation}
S=\frac{k}{4\pi} \int_{D \times {\bf R}^1} a da,~~ a = a_\mu dx^\mu,
{}~~ada \equiv a\wedge da
\end{equation}
where $a_{\mu}$ is a real field.

The action S is invariant under diffeos of the solid cylinder and does not
permit a natural choice of a time function.  As time is all the same
indispensable in the canonical approach, we arbitrarily choose a time function
denoted henceforth by $x^0$. Any constant $x^0$ slice of the solid
cylinder is then the disc $D$ with
coordinates $x^1$, $x^2$.

It is well known that the phase space of the action $S$ is
described by the equal
time Poisson brackets (PB's)
\be
\left \{a_{i}(x),a_{j}(y)\right \}=\epsilon_{ij}\frac{2\pi}{k}\delta^{2}(x-y)
 ~~ {\rm for}~ i,j=1,2 ,
\;\;\;\;\epsilon_{12}=-\epsilon_{21}=1
\ee
and the constraint
\be
\partial_{i}a_{j}(x) - \partial_{j}a_{i}(x) \equiv f_{ij}(x) \approx 0
\ee
where $\approx$ denotes weak equality in the sense of Dirac$^{(9)}$.
All fields
are evaluated at the same time $x^0$ in these equations, and this will continue
to be the case when dealing with the canonical formalism or quantum operators
in the remainder of the paper.  The connection $a_0$ does not occur as a
coordinate of this phase space.  This is because, just as in electrodynamics,
its conjugate momentum is weakly zero and first class and hence eliminates
$a_0$ as an observable.

The constraint (4.3) is somewhat loosely stated.  It is important to formulate
it more accurately by first smearing it
with a suitable class of ``test" functions
$\Lambda^{(0)}$.  Thus we write,
instead of (4.3),
\be
g(\Lambda^{(0)}) : \; =\frac{k}{2\pi}
 \int_{D} \Lambda^{(0)}(x) da(x) \approx 0 \; .
\ee
It remains to state the space ${\cal T}^{(0)}$ of test functions
$\Lambda^{(0)}$
{}.
For this purpose, we recall that a functional on phase space can be relied on
to generate well defined canonical transformations only if it is
differentiable.  The meaning and implications of this remark can be illustrated
here by varying $g(\Lambda^{(0)})$ with respect to $a_{\mu}$:
\be
\delta g (\Lambda^{(0)}) =
\frac{k}{2\pi}\left ( \int_{\partial D}\Lambda^{(0)} \delta a -
\int_{D}d \Lambda^{(0)} \delta a \right ).
\ee
By definition, $g(\Lambda^{(0)})$ is differentiable in $a$ only if the boundary
term
-- the first term -- in (4.5) is zero.  We do not wish to constrain the phase
space by legislating $\delta a$ itself to be zero on $\partial D$ to achieve
this goal.  This is because we have a vital interest in regarding fluctuations
of $a$ on $\partial D$ as dynamical and hence allowing canonical
transformations
which change boundary values of $a$.  We are thus led to the following
condition
on functions $\Lambda^{(0)}$ in ${\cal T}^{(0)}$:
\be
\Lambda^{(0)} \mid_{\partial D} = 0 \;.
\ee

 It is useful to illustrate the sort of troubles we will encounter if (4.6) is
dropped.  Consider

\be
q(\Lambda) =  \frac { k } {2\pi}
\int_{D} d\Lambda a
\ee
It is perfectly differentiable in $a$ even if the function $\Lambda$ is nonzero
on $\partial D$.  It creates fluctuations
$$\delta a\mid_{\partial D} = d \Lambda \mid_{\partial D}$$  of $a$ on
 $\partial D$ by canonical
transformations.  It is a function we wish to admit in our canonical approach.
Now consider its PB with $g(\Lambda^{(0)})$:
\be
\{g(\Lambda^{0}), q(\Lambda)\} = \frac{k}{2\pi} \int d^{2}x d^{2}y
\Lambda^{(0)}(x)\epsilon^{ij}\left [\partial_{j}\Lambda (y)\right ]
\left [\frac{\partial}
{\partial
x^{i}} \delta^{2}(x-y)\right ]
\ee
where $\epsilon^{ij} = \epsilon_{ij}$. This expression is quite ill
defined if $$\Lambda^{(0)}\mid_{{\partial D}}\neq 0.$$ Thus integration on $y$
first gives zero for (4.8).  But if we integrate on $x$  first,
treating derivatives of distributions by usual rules, one finds instead,
\be
- \int_{D} d\Lambda^{0}d\Lambda = -\int_{\partial D}\Lambda^{0}d\Lambda\ .
\ee
Thus consistency requires the condition (4.6).

We recall that a similar situation occurs in QED.  There, if $E_j$ is the
electric field, which is the momentum conjugate to the potential $a_j$, and
$j_0$ is the charge density, the Gauss law can be written as
\be
\bar {g}(\bar{\Lambda}^{(0)}) = \int d^{3}x
\bar{\Lambda}^{(0)}(x)\left[\partial
_{i}
E_{i}(x)-j_{0}(x)\right]\approx 0\,.
\ee

\noindent
Since
\be
\delta \bar{g} (\bar{\Lambda}^{(0)})  = \int_{r=\infty} r^{2} d\Omega
\bar{\Lambda}^{(0)}(x) \hat{x}_{i}\delta E_{i}
-\int d^{3}x \partial_{i}\bar {\Lambda}^{(0)}(x)\delta E_i(x),
r=\mid\vec{x}\mid, \hat{x} = \frac {\vec{x}}{r}
\ee
for the variation $\delta E_i$ of $E_i$, differentiability requires
\be
\bar{\Lambda}^{(0)}(x)\mid_{r=\infty}=0.
\ee
$[d\Omega$ in (4.11) is the usual volume form of the two sphere ].
The charge, or equivalently the generator of the global U(1)
transformations, incidentally is the analogue of $q(\Lambda)$.  It is got by
partial integration on the first term.  Thus let
\be
\bar{q}(\bar{\Lambda}) = -\int d^3x\partial_{i}\bar{\Lambda}(x) E_{i}(x)
-\int d^{3}x\bar\Lambda(x)j_{0}(x) \,.
\ee
This is differentiable in $E_i$ even if $\bar {\Lambda} \mid_{r=\infty}  \neq
0$
 and
generates the gauge transformation for the gauge group element
$e^{i\bar{\Lambda}}$. It need not to vanish on quantum states if
$\bar{\Lambda}\mid_{r=\infty}\neq 0$, unlike $\bar{g}(\bar{\Lambda}^{(0)})$
which is associated with the Gauss law $\bar
{g}(\bar{\Lambda}^{(0)}) \approx 0$.  But if $\bar{\Lambda}\mid_{r=\infty}=0$,
 it becomes
the Gauss law on partial integration and annihilates all physical states.  It
follows that if $(\bar{\Lambda}_{1}-\bar{\Lambda}_{2}) \mid_{r=\infty} =0$,
then
$\bar {q} (\bar{\Lambda}_1) = \bar{q} (\bar {\Lambda}_{2})$ on physical states
which are thus sensitive only to the boundary values of test functions.  The
nature of their response determines their charge.  The conventional electric
charge of QED is $\bar {q}({\bf \bar{ 1}})$ where $\bar {\bf 1}$ is the
constant
function with value  1.

The constraints $g(\Lambda^{(0)})$ are first class since
\begin{eqnarray}
\left \{g(\Lambda_{1}^{(0)}), g(\Lambda_{2}^{(0)}) \right \}
&=& \frac{k}{2\pi} \int_{D} d\Lambda_{1}^{(0)} d \Lambda_{2}^{(0)} \nonumber \\
&=& \frac {k}{2\pi} \int_{\partial D} \Lambda_{1}^{(0)} d\Lambda_{2}^{(0)}
\nonumber \\
&=& 0 \;\;\; {\rm for}~ \Lambda_{1}^{(0)},~\Lambda_{2}^{(0)} \in\;
{\cal T}^{(0)}\;.
\end{eqnarray}
$g(\Lambda^{(0)})$ generates the gauge transformation
$a \rightarrow  a+d\Lambda^{(0)}$
of $a$.

Next consider $q(\Lambda)$ where $\Lambda\mid_{\partial D}$ is not necessarily
zero.  Since

\begin{eqnarray}
\left \{q(\Lambda),g(\Lambda^{(0)}) \right \} &=&
-\frac{k}{2\pi}\int_{D} d\Lambda d\Lambda^{(0)} \nonumber \\
&=&\frac{k}{2\pi}  \int_{\partial D} \Lambda^{(0)} d\Lambda =0 \;\;
{\rm for}~\Lambda^{(0)} \in {\cal T}^{(0)},
\end{eqnarray}
they are first class or the observables of the theory.  More precisely
observables are obtained after identifying $q(\Lambda_{1})$
with $q(\Lambda_{2})$ if
$(\Lambda_{1}-\Lambda_{2}) \in {\cal T}
^{(0)}$.  For then,
$$q(\Lambda_{1})-q(\Lambda_{2}) = - g(\Lambda_{1} -\Lambda_{2})
\approx 0.$$  The functions $q(\Lambda)$
generate gauge transformations $a\rightarrow a+d\Lambda$ which
do not necessarily vanish
on $\partial D$.

It may be remarked that the expression for $q(\Lambda)$ is obtained from
$g(\Lambda^{(0)})$ after a partial integration and a subsequent substitution of
$\Lambda$ for $\Lambda^{(0)}$.  It too  generates gauge transformations like
$g(\Lambda^{(0)})$, but the test function space for the two are different.  The
pair $q(\Lambda),g(\Lambda^{(0)})$ thus resemble the pair
$\bar{q}({\bar\Lambda}),\bar{g}(\bar{\Lambda}^{(0)})$ in QED.
The resemblance suggests
that we think of $q(\Lambda)$ as akin to the generator of a global symmetry
transformation. It is natural to do so for another reason as well:
the Hamiltonian is a constraint for a first order Lagrangian such as the one we
have here, and for this Hamiltonian, $q(\Lambda)$ is a constant of motion.

In quantum gravity, for asymptotically flat spatial slices, it is often the
practice to include a surface term in the Hamiltonian which would otherwise
have been a constraint and led to
trivial evolution.  However, we know
of no natural choice of such a surface term, except zero, for the CS theory.

The PB's of $q(\Lambda)$'s are easy to compute:
\be
\{q(\Lambda_{1}),q(\Lambda_{2})\} = \frac {k}{2\pi} \int_{D} d\Lambda_{1}
d\Lambda_{2} =
\frac {k}{2\pi} \int_{\partial D} \Lambda_{1} d \Lambda_{2} \;.
\ee
Remembering that the observables are characterized by boundary values of test
functions, (4.16) shows that the observables generate a U(1) Kac-Moody
algebra$^{(8)}$ localized on $\partial D$. It is a Kac-Moody algebra for ``zero
momentum'' or ``charge''. For if $\Lambda \mid_{\partial D}$ is  a constant,
it can be extended as a constant
function to all of $D$ and then $q(\Lambda) = 0$.  The central charges and
hence
the representation of (4.16) are different for $k>0$ and $k<0$, a fact which
reflects parity violation by the action $S$.

Let $\theta$ (mod $2\pi$) be the coordinate on $\partial D$ and $\phi$
a free massless
scalar field moving with speed $v$ on $\partial D$ and obeying the equal time
PB's
\be
\{\phi(\theta), \dot {\phi}(\theta^\prime)\}=\delta(\theta-\theta^\prime)\;.
\ee
If $\mu_{i}$ are test functions on $\partial D$ and
$\partial_{\pm}=\partial_{x^{0}}\pm v \partial_{\theta}$,
then
\be
\left \{\frac{1}{v} \int \mu_{1} (\theta)\partial_{\pm}
\phi(\theta),\frac{1}{v}\int
\mu_{2}(\theta)\partial_{\pm}\phi(\theta) \right \}=\pm2 \int \mu_{1}(\theta)
d \mu_{2}(\theta) ,
\ee
the remaining PB's being zero. Also $\partial_{\mp}\partial_{\pm}\phi = 0$.
Thus
the algebra of observables is isomorphic to that generated by the left moving
$\partial_{+} \phi$ or the right moving $\partial_{-}\phi$.

The CS interaction is invariant under diffeos of $D$.  An infinitesimal
generator of a diffeo with vector field $V^{(0)}$ is$^{(10)}$
\be
\delta(V^{(0)})= -\frac {k}{2\pi} \int_{D}V^{(0)i}a_{i}da.
\ee
The differentiability of $\delta(V^{(0)})$ imposes the constraint
\be
V^{(0)}\mid_{\partial D}=0\;.
\ee
Hence, in view of (4.4) as well, we have the result
\be
\delta(V^{(0)})=-\frac{k}{4\pi}\int_{D}a{\cal L}_{V^{(0)}}a \approx 0\;.
\ee
where ${\cal L}_{V^{(0)}}a$ denotes the Lie derivative
of the one form $a$ with respect to  the vector field $V^{(0)}$ and is given
by $$({\cal L}_{V^{(0)}}a)_{i}
= \partial_{j}a_{i}V^{(0)j} + a_{j}\partial_{i}V^{(0)j}.$$

Next, suppose that $V$ is a vector field on $D$ which on $\partial D$ is
tangent
 to
$\partial D$,
\be
V^{i}\mid_{\partial D}(\theta) = \epsilon (\theta)
\left( \frac {\partial x^i}{\partial\theta} \right ) \mid_{\partial D} ,
\ee
$\epsilon$ being any function on $\partial D$ and $x^{i}\mid_{\partial D}$ the
restriction of $x^{i}$ to $\partial D$. $ V$ thus generates a diffeo
mapping $\partial D$ to $\partial D$. Consider next
\begin{eqnarray}
l(V)&=&\frac{k}{2\pi}\left(~\frac{1}{2}\int_{D} d(V^{i}a_{i}a) - \int_{D}
V^{i}a_{i}da \right)~ \nonumber \\
&=&-\frac{k}{4\pi}\int_{D}a{\cal L}_{V}a \ .
\end{eqnarray}
Simple calculations show that $l(V)$ is differentiable in $a$ even if
$\epsilon(\theta) \neq 0$ and generates the infinitesimal diffeo associated
with the vector field $V$. We show in
subsection 4.2 that it is, in fact, related to $q(\Lambda)$'s by the Sugawara
construction.

The expression (4.23) for the diffeo generators of observables given in the
first paper of ref. 6 seems to be new.

As final points of this subsection, note that
\be
\{l(V),g(\Lambda^{(0)})\}= g(V^{i}\partial_{i}\Lambda^{(0)})
=g({\cal L}_{V}\Lambda^{(0)})\approx 0\;,
\ee
\be
\{l(V),q(\Lambda)\}=q(V^{i}\partial_{i}\Lambda)=q({\cal L}_{V}\Lambda),
\ee
\be
\{l(V),l(W)\} =l({\cal L}_{V}W)
\ee
where ${\cal L}_{V}W$ denotes
the Lie derivative of the vector field $W$ with respect to the vector
field $V$ and is given
by $$({\cal L}_{V}W)^{i} =  V^{j}\partial_{j}W^{i}-W^{j}\partial_{j}V^{i}.$$
$l(V)$ are first class in view of (4.24).
Further, after the imposition of constraints, they are entirely characterized
by $\epsilon(\theta)$, the equivalence class of $l(V)$ with the same
$\epsilon(\theta)$ defining an observable.

\vskip .3in
\centerline {\bf 4.2 QUANTIZATION}
\vskip .3in

\indent
Our strategy for quantization relies on the observation that if
$$\Lambda\mid_{\partial D}(\theta)=e^{iN\theta} , $$ then the PB's (4.16)
become those of
creation and annihilation operators.  These latter can be identified with the
similar operators of the chiral fields $\partial_{\pm}\phi$

Thus let $\Lambda_{N}$ be any function on  $D$ with boundary value
$e^{iN\theta}$:
\be
\Lambda_{N}\mid_{\partial D}(\theta)= e^{iN\theta},\;\;N \in {\bf Z}\;.
\ee
These $\Lambda_{N}$'s exist.  All $q(\Lambda_{N})$ with the same
$\Lambda_{N}\mid_{\partial D}$ are weakly equal and define the same observable.
Let $\langle q(\Lambda_{N})\rangle$ be this equivalence class
and $q_{N}$ any member
thereof. [$q_{N}$ can also be regarded as the equivalence class itself.] Their
PB's follow from (4.16):
\be
\{q_{N},q_{M}\} = - i N k \delta_{N+M,0} \;.
\ee
The $q_{N}$'s are the CS constructions of the Fourier modes of a massless
chiral scalar field on the circle $S^{1}$.

The CS construction of the diffeo generators $l_{N}$ on $\partial D$ (the
classical analogues of the Virasoro generators) are similar.  Thus let

 $$< l (V_{N})>$$ be the equivalence class of $l (V_{N})$
defined by the constraint
\be
V^{i}_{N} \mid_{\partial D} = e^{iN\theta}~
\left (\frac {\partial x^i}{\partial \theta}\right)\mid_{\partial D},
\sp N \in {\bf Z},
\ee
$(x^{1},~ x^{2})\mid_{\partial D}(\theta)$ being chosen to be
$R(\cos \theta,\sin
\theta)$ where $R$ is the radius of $D$.
Let $l_N$ be any member of $$<l(V_N)>~.$$ It can be
verified that
\be
\{l_N,q_M\} = i M q_{N+M} \;,
\ee
\be
\{l_N,l_M\} = - i(N-M)~ l_{N+M} \;.
\ee
These PB's are independent of the choice of the representatives from their
respective equivalence classes.  Equations (4.28), (4.30) and (4.31) define the
semidirect product of the Kac-Moody algebra and the Witt algebra (Virasoro
algebra without the central term) in its classical version.

We next show that
\be
l_N \approx \frac {1}{2k} \sum_{M}q_{M}~q_{N-M}
\ee
which is the classical version of the Sugawara construction$^{(8)}$ .

For convenience, let us introduce polar coordinates $r,\theta$ on $D$ ( with
$r = R$ on $\partial D$ ) and write the fields and test functions
as functions of polar coordinates. It is then clear that
\be
l_N \equiv l(V_N) = \frac{k}{4 \pi}\int_{\partial D}d\theta e^{iN\theta}
a^{2}_{\theta}(R,\theta) - \frac{k}{2\pi}
\int_{D}V^{l}_{N}(r,\theta)a_{l}(r,\theta)da(r,\theta)
\ee
where $ a = a_{r}dr + a_{\theta}d\theta.$

Let us next make the choice
\be
 e^{iM\theta}\lambda (r),~ \lambda(0)=0 \;,~~\lambda(R)=1
\ee
for $\Lambda_{M}$.
Then
\be
q_M = q(e^{iM\theta}\lambda(r)).\ee
Integrating (4.35) by parts we get
\be
q_M = \frac {k}{2 \pi}\left (
\int_{\partial D} d\theta e^{iM\theta}a_{\theta}(R,\theta)
- \int_{D} dr d{\theta} \lambda (r) e^{iM\theta} f_{r \theta}(r,\theta) \right
)
\ee
\noindent where  $f_{r \theta}$ is defined by $da = f_{r\theta}dr \wedge
d\theta$. Therefore
\begin{eqnarray}
\frac{1}{2k} \sum_{M} q_{M}q_{N-M}~=
&+&\frac{k}{4\pi}\int_{\partial D} d\theta  e^{iN\theta} a^2_{\theta}(R,\theta)
 \nonumber \\
&-&\frac{k}{2 \pi}\int_{D} dr d\theta  e^{iN\theta}\lambda (r)
 a_{\theta}(R, \theta) f_{r \theta}(r, \theta)
\nonumber \\
&+&\frac{k}{4 \pi}\int_{D} dr d\theta dr^\prime \lambda (r) \lambda
(r^{\prime})e^{iN\theta}
f_{r \theta}(r, \theta) f_{r \theta}(r^{\prime}, \theta)
\end{eqnarray}
where the completeness relation
$$\sum_N e^{iN(\theta-\theta^\prime)} = 2\pi \delta (\theta-\theta^\prime)$$
has been used.

The test functions for the Gauss law in the last term in (4.37)
involves $f_{r \theta}$ itself. We therefore interpret it to be zero and get
\be
\frac{1}{2k} \sum_{M} q_{M}q_{N-M} \approx
 \frac{k}{4\pi} \int_{\partial D} a^2_{\theta}(R,\theta) e^{iN\theta}d
\theta - \frac{k}{2\pi}\int_{D}dr d\theta
e^{iN\theta}\lambda(r)a_{\theta}(R,\theta)f_{r \theta}(r,\theta).
\ee

 Now in view of  (4.29) and (4.34), it is clear that
\be
V^{l}_{N}(r,\theta)a_{l}(r,\theta) - e^{iN\theta}\lambda(r)a_{\theta}(R,\theta)
= 0 ~~~ {\rm on}~~ \partial D.
\ee
Therefore $$l_N \approx \frac {1}{2k} \sum_{M} q_{M}q_{N-M} $$
which proves (4.32).

We can now proceed to quantum field theory. Let ${\cal
G}(\Lambda^{(0)}),Q(\Lambda_N),Q_N$ and $L_N$ denote the quantum operators for
$g(\Lambda^{(0)}), q(\Lambda_N),q_N$ and $l_N$.  We then impose the constraints
\be
{\cal G}(\Lambda^{(0)}) | \cdot \rangle = 0
\ee
on all quantum states.  It is an expression of their gauge invariance.  Because
of this equation, $Q(\Lambda_N)$ and $Q(\Lambda ^\prime_N)$ have the same
action
on the states if $\Lambda_N$ and $\Lambda ^\prime_N$ have the same boundary
values.  We can hence write
\be
Q_N | \cdot \rangle = Q(\Lambda_N) | \cdot \rangle \;.
\ee
Here, in view of (4.28), the commutator brackets of $Q_N$ are
\be
[Q_N,Q_M] = N k \delta_{N+M,0}\;.
\ee

Thus if $k>0 \;\;(k<0),Q_N$ for $N>0 ~ (N<0)$ are annihilation operators
( upto a normalization ) and
$Q_{-N}$ creation operators.  The ``vacuum'' $| 0 >$ can therefore be defined
by
\be
Q_N \mid 0> = 0 \;\;{\rm if}~ Nk>0\;.
\ee
The excitations are got by applying $Q_{-N}$ to the vacuum.

The quantum Virasoro generators are the normal ordered forms of their classical
expression$^{(8)}$  :
\be
L_N =  \frac {1}{2k}:~\sum_{M}Q_M Q_{N-M}:
\ee
They generate the Virasoro algebra for central charge $c=1$ ~:
\be
\left [L_N,L_M \right ] = (N-M)L_{N+M}
+ \frac {c}{12} (N^3-N) \delta_{N+M,0}~,~c=1 \;\;.
\ee

When the spatial slice is a disc, the observables are all given by $Q_N$ and
our quantization is complete.  When it is not simply connected, however, there
are further observables associated with the holonomies of the connection $a$
and
they affect quantization.  We will not examine quantization for nonsimply
connected spatial slices here.

The CS interaction does not fix the speed $v$ of the scalar field in (4.18) and
so its Hamiltonian, a point previously emphasized by
Fr\"{o}hlich and Kerler,  and Fr\"{o}hlich and Zee$^{(4)}$ .
This is but reasonable.  For if we could fix $v$, the Hamiltonian H for $\phi$
could naturally be taken to be the one for a free massless chiral scalar field
moving with speed $v$.  It could then be used to evolve the CS observables
using
the correspondence of this field and the former.  But we have seen that no
natural nonzero Hamiltonian exists for the CS system.  It is thus satisfying
that we can not fix $v$ and hence a nonzero H.

In the context of Fractional Quantum Hall Effect, as we have seen in Section 3,
the following generalization
of the CS action has become of interest$^{(3)}$ :
\be
S ^\prime = \frac{k}{4\pi} {\cal K}_{IJ}
\int_{D \times {\bf R}^{1}}a^{(I)} da^{(J)}.
\ee
Here the sum on $I,J$ is from $1$ to $m$, $a^{(I)}$ is associated with
the current $j^{(I)}$ in the $I^{th}$ Landau level and
${\cal K}$ is a certain invertible symmetric
real $F \times F$ matrix . By way of further illustration
of our approach to
quantization,  we now outline the quantization of (4.46) on $D \times {\bf
R}^{1}$.

The phase space of (4.46) is described by the PB's
\be
\left \{ a^{(I)}_{i} (x), a^{(J)}_j(y) \right \} = \epsilon_{ij}\frac{2\pi}{k}
{\cal K}^{-1}_{IJ}\delta^{2}(x-y),~~x^0=y^0
\ee
and the first class constraints
\be
g^{(I)}(\Lambda^{(0)}) =
\frac{k}{2\pi} \int_D \Lambda^{(0)}da^{(I)}\approx 0\ , ~~ \Lambda^{(0)} \in
{\cal T}^{(0)}
\;.
\ee
with zero PB's.

The observables are obtained from the first class variables
\be
q^{(I)}(\Lambda) = \frac{k}{2\pi} \int_D d\Lambda a^{(I)}
\ee
after identifying $q^{(I)}(\Lambda)$ with $q^{(I)}(\Lambda^\prime)$ if
$(\Lambda-\Lambda^\prime)\mid_{\partial D}=0$. The PB's of $q^{(I)}$'s are
\be
\left \{q^{(I)}(\Lambda^{(I)}_{1}), q^{(J)}(\Lambda^{(J)}_{2})\right \} =
\frac{k}{2\pi}{\cal K}^{-1}_{IJ} \int_{\partial D}
\Lambda^{(I)}_{1}d\Lambda^{(J)}_{2}.
\ee

Choose a $\Lambda^{(I)}_{N}$ by the requirement $\Lambda^{(I)}_N \mid_{\partial
D}(\theta)= e^{iN\theta}$ and let $q^{(I)}_N$ be any member of the equivalence
class $<q^{(I)}(\Lambda^{(I)}_N)>$ characterized by such $\Lambda^{(I)}_N$.
Then
\be
\left \{q^{(I)}_N,q^{(J)}_M \right \}= -i {\cal K}^{-1}_{IJ} Nk\delta_{N+M,0}~.
\ee

As ${\cal K}^{-1}_{IJ}$ is real symmetric, it can be diagonalized by a real
orthogonal
transformation $M$ and has real eigenvalues $\lambda_{\rho} ~(\rho=1,2,...,m)$.
As ${\cal K}^{-1}_{IJ}$ is invertible, $\lambda_{\rho} \neq 0$. Setting
\be
q_N(\rho)=M_{\rho I}q^{(I)}_N
\ee
we have
\be
\left \{q_N(\rho), q_M(\sigma)\right \} = -i \lambda_\rho Nk
\delta_{\rho\sigma}
\delta_{N+M,0}~~.
\ee
(4.53) is readily quantized. If $Q_N(\rho)$ is the quantum operator for
$q_N(\rho)$,
\be
\left [Q_N(\rho),Q_M(\sigma)\right] =  \lambda_{\rho} Nk
\delta_{\rho\sigma}\delta_{N+M,0}~.
\ee
(4.54) describes $m$  harmonic oscillators or edge currents.  Their chirality,
or the chirality of the corresponding massless scalar fields, is governed by
the
sign of $\lambda_\rho$.

The classical diffeo generators for the independent oscillators $q_N(\rho)$ and
their quantum versions can be written down using the foregoing discussion.  The
latter form $m$ commuting Virasoro algebras, all for central charge 1.

\vskip .5in
\centerline {\bf ACKNOWLEDGEMENTS}
\vskip .3in

 A.P.B. would like to thank Diptiman Sen for useful discussions, especially
regarding the edge currents in a Quantum Hall system. The work of A.P.B. was
supported by the U.S. Department of Energy under contract number
DE-FG02-85ER40231. The work of A.M.S. was supported by the Theoretical Physics
Institute at the University of Minnesota and by the U.S. Department of Energy
under  contract number DE-AC02-83ER40105.

\newpage

\centerline{$\underline{\it REFERENCES}$}

 No attempt will be made here to compile an adequate bibliography.
The following books and papers may be consulted for further guide to
literature.

\begin{enumerate}

\item For reviews of QHE, see, ``The Quantum Hall Effect", edited by
R.E. Prange and S. Girvin (Springer-Verlag, New York, 1987);
G. Morandi, ``Quantum Hall Effect" (Bibliopolis, Napoli, 1988);
A.P. Balachandran, E. Ercolessi, G. Morandi and A.M. Srivastava, `` The
Hubbard Model and Anyon Superconductivity", Int. J. Mod. Phys. B4 (1990) 2057
and Lecture Notes in Physics, Volume 38 (World Scientific, Singapore, 1990);
F. Wilczek, ``Fractional Statistics and Anyon Superconductivity" (World
Scientific, Singapore, 1990) and articles therein. See also ref. 2.

\item T. Chakraborty and P. Pietil\"{a}inen, ``The Fractional
Quantum Hall Effect", (Springer-Verlag, Berlin, 1988).

\item S.C. Jhang, T.H. Hansson and S. Kivelson, Phys. Rev. Lett.
62 (1989) 82; B. Blok and X.G. Wen, Institute for Advanced Study
preprint IASSNS -HEP - 90/23 (1990).

\item J. Fr\"{o}hlich and T. Kerler, Nucl. Phys. B 354 (1991) 369;
J. Fr\"{o}hlich and A. Zee, Institute for Theoretical Physics,
Santa Barbara preprint  NSF-ITP-91-31 (1991).

\item B.I. Halperin, Phys. Rev. B25 (1982) 2185.

\item A.P. Balachandran, G. Bimonte, K.S. Gupta and A. Stern,
Syracuse University preprints SU-4228-477 and 487 (1991).

\item E. Witten, Commun. Math. Phys. 121 (1989) 351.

\item For a review, see P. Goddard and D. Olive, Int. J. Mod. Phys.
A1 (1986) 303.

\item Cf. A.P. Balachandran, G. Marmo, B.-S. Skagerstam and A. Stern,
``Classical Topology and Quantum States" (World Scientific,
Singapore, 1991), Part I.

\item E. Witten, Nucl. Phys. B311 (1988) 46.

\end{enumerate}

\newpage

\centerline {\bf FIGURE CAPTIONS}

Figure 1 : Electrons are confined in the annulur region $r_1 < r < r_2$
in the 1-2 plane with a uniform magnetic field $B_0$ directed along the
third axis in that region. Additional magnetic flux $\Phi$ is confined
to the region $r < r_1$.

Figure 2 : Shift of the energy levels at the edges $r = r_1$ and
$r = r_2$ due to the boundary effects.

\end{document}